\documentclass[5p,twocolumn,times]{elsarticle}
\journal{Physics Letters B}

\usepackage{graphicx}
\usepackage{amsmath}
\usepackage{amssymb}
\usepackage{tikz}
\usepackage{epstopdf}
\usepackage{dcolumn}
\usepackage[percent]{overpic}  
\usepackage{bm}
\usepackage{xcolor,xspace}
\usepackage{braket}
\usepackage{wasysym}  
\usepackage[ulem=normalem]{changes}
\usepackage{hyperref}
\hypersetup{colorlinks=True,urlcolor=blue,linkcolor=blue,citecolor=blue,filecolor=black}

\colorlet{Changes@Color}{red}  

\newcommand{\ba}{\begin{eqnarray}}
\newcommand{\ea}{\end{eqnarray}}
\newcommand{\ban}{\begin{eqnarray*}}
\newcommand{\ean}{\end{eqnarray*}}

\newcommand{\bsub}{\begin{subequations}}
\newcommand{\esub}{\end{subequations}}

\def\ket#1{|#1\rangle}

\def\bsu3{\overline{{\rm SU(3)}}}
\def\bso6{\overline{{\rm SO(6)}}}
\def\bPi2{\overline{\Pi}^{(2)}}

\def\b0{\beta_0}
\def\beq{\beta_{\rm eq}}
\def\g0{\gamma_0}

\def\bsu3nu{\overline{{\rm SU_{\nu}(3)}}}

\begin{document}

\title{Geometry of Configuration Mixing in Bose-Fermi Systems}

\author[]{A. Leviatan\corref{cor1}}
\ead{ami@phys.huji.ac.il}

\author[]{N. Gavrielov}
\ead{noam.gavrielov@mail.huji.ac.il}

\cortext[cor1]{Corresponding author}

\address{Racah Institute of Physics,
  The Hebrew University, Jerusalem 91904, Israel}

\date{\today}

\begin{abstract}
  A geometric interpretation for an algebraic interacting
  boson-fermion model with
  configuration mixing is presented.
The formalism is based on an extended Bose-Fermi matrix
coherent states and is applied to gain insight on
intertwined quantum shape-phase transitions and
  shape coexistence in odd-mass Nb nuclei.
  \end{abstract}

\begin{keyword}
Bose-Fermi systems;
Algebraic models;
Coherent states;
Interacting boson-fermion model;
Quantum shape-phase transitions;
Shape coexistence in nuclei
\end{keyword}

\maketitle

Algebraic models
provide a convenient framework for tractable yet
detailed calculations of observables
and for unraveling global trends of structure and symmetries
in diverse dynamical
systems~\cite{IBMBook,IBFMBook,VibronBook,FrankIsaBook}.
A key advantage is that these models
are amenable for both quantum
and classical analysis.
Given a boson system with a prescribed algebraic structure
and Hilbert space ${\cal H}_{\rm b}$, a geometric
visualization is obtained
by an expectation value of the Hamiltonian in
a coherent state
$\Phi_{\rm b}(y)\in {\cal H}_{\rm b}$,
which depends on classical (geometric)
variables $y$, parameterizing
the relevant coset space~\cite{Gilmore79}.
For two (or more) configurations,
the Hilbert space is a direct sum
${\cal H}_{\rm b;1} \oplus{\cal H}_{\rm b;2}$,
where ${\cal H}_{\rm b;i}$ correspond to orthogonal subspaces.
Geometry is introduced
via a matrix $\Omega(y)$
of the Hamiltonian in a set of coherent states,
$\Phi_{\rm b; i}(y)\in {\cal H}_{\rm b;i}$.
Notable examples~of such constructions are known
for the interacting boson model (IBM)~\cite{IBMBook},
based on a $U(6)$ algebra,
and for its extension to include configuration mixing
(IBM-CM)~\cite{Duval1981,Duval1982}, in studies of
the classical limit~\cite{GinoKir80,Diep80,Isa81},
shape evolution~\cite{DiepSchol80,FengGilmDean81,Cejnar09,Cejnar10}
and shape
coexistence~\cite{Vargas04,FrankIsaIac06,Morales08,Ramos14,Nomura16}
in even-even~nuclei.

For a Bose-Fermi system,
the Hilbert space is a  tensor product of the bosonic
and fermionic spaces
${\cal H}_b\otimes{\cal H}_f$, and the matrix $\Omega(y)$
is in product states
$\Phi_{\rm b}(y)\phi_{{\rm f};m}$, 
where $\phi_{{\rm f};m}\in {\cal H}_f$ and
$m$ enumerates fermion basis states.
This procedure was implemented in the interacting
boson-fermion model (IBFM)~\cite{IBFMBook},
furnishing a formalism to analyze 
its geometric
properties~\cite{Lev88,LevShao89,Alonso92}
and shape-phase
transitions~\cite{Boyukata2010,Petrellis2011,Boyukata2021}
relevant to odd-even nuclei.
In the present Letter, we
expand the formalism to encompass the geometry
of a Bose-Fermi system with multiple configurations
for which the
Hilbert space is of the form
$\left [{\cal H}_{b;1}\oplus{\cal H}_{b;2}\right ]
\otimes {\cal H}_f$. For that, we introduce
an extended matrix coherent states and
apply it to the recently introduced
interacting boson-fermion model
with configuration mixing
(IBFM-CM)~\cite{GavLevIac22b,Gav23}
to describe the evolving geometry and coexistence
in odd-mass Nb isotopes. This construction is
motivated by the recent advances in experimental
studies of shape coexistence in
nuclei~\cite{Garrett22}, which
underscore the importance of configuration mixing
for the interpretation of this
ubiquitous phenomena.

Odd-A nuclei are treated in the interacting boson-fermion
model (IBFM)~\cite{IBFMBook}, as a system of monopole ($s$) 
and quadruple ($d$) bosons, representing valence nucleon 
pairs, and a single (unpaired) nucleon.
We focus the discussion to an odd fermion in 
a single-$j$ orbit, for which the relevant algebra is
$U_{\rm b}(6)\otimes U_{\rm f}(2j+1)$.
The (single-configuration) IBFM Hamiltonian has the form
\ba
\hat H(\xi) = \hat H_{\rm b} + \hat H_{\rm f}
+ \hat V_{\rm bf} ~.
\label{Hibfm}
\ea
The boson part is
$ \hat H_{\rm b}(\epsilon_d,\kappa,\chi) =
  \epsilon_d\,
  \hat n_d + \kappa\,
  \hat Q_\chi \cdot \hat Q_\chi$,
  where $\hat{n}_d\!=\!\sum_{m}d^{\dag}_md_m$,
  $\hat Q_\chi \!=\!
d^\dag s+s^\dag \tilde d +\
\chi\, (d^\dag \tilde d)^{(2)}$ and
$\tilde{d}_m\!=\!(-1)^{m}d_{-m}$.
The fermion part involves the
number operator, 
$\hat{H}_{\rm f} = \epsilon_{j}\,\hat{n}_j$,
and the boson-fermion part is composed of monopole,
quadrupole and exchange  terms,
$\hat{V}_{\rm bf}(\chi,A,\Gamma,\Lambda) =
A\,\hat{n}_d\,\hat{n}_j + \Gamma\,\hat{Q}_{\chi}\cdot
( a_{j}^{\dag }\, \tilde{a}_{j} )^{(2)}
+ \Lambda
\sqrt{2j+1}:[ ( d^{\dag }\, \tilde{a}_{j})^{(j)}\times
  ( \tilde{d}\, a_{j}^{\dag })^{(j)}]^{(0)}:$,
where
$:\!-\!:$ denotes normal ordering
and $\tilde{a}_{j,m} \!=\! (-1)^{j-m}a_{j,-m}$.
The parameters of $\hat{H}(\xi)$ are denoted by
$\xi=(\epsilon_d,\kappa,\chi,\epsilon_j,A,\Gamma,\Lambda$).

Geometry is encoded in
a matrix potential of the
Hamiltonian~(\ref{Hibfm})
in the following basis~\cite{Lev88,LevShao89,Alonso92},
\ba
\ket{j,m;\beta,\gamma;N} =
\ket{j,m}\otimes\ket{\beta,\gamma;N} ~,
\ea
which involves the product of a $j$-fermion
$\ket{j,m}= a^{\dag}_{jm}\ket{0}$
and a projective coherent state
of $N$ bosons~\cite{GinoKir80,Diep80},
\ba
\ket{\beta,\gamma;N} = (N!)^{-1/2}(b^\dag_c)^N\ket{0} ~.
\ea
Here
$b^\dag_c \!=\! \tfrac{1}{\sqrt{2\beta+1}}
[\beta\cos\gamma d^\dag_0 +
  \tfrac{1}{\sqrt{2}}\beta\sin\gamma
  (d^\dag_2 + d^\dag_{-2}) + s^\dag]$ and
$(\beta,\gamma)$ are quadrupole shape variables.
The resulting matrix, $\Omega_{N}(\beta,\gamma;\xi)$,
is real and symmetric with entries 
$E_{\rm b}(\beta,\gamma;N)\delta_{m_1,m_2} +
\epsilon_j\,\delta_{m_1,m_2} + Ng_{m_1,m_2}(\beta,\gamma)$.
Here $E_{\rm b}(\beta,\gamma;N)$ is the expectation value
of $\hat{H}_{\rm b}$ in $\ket{\beta,\gamma;N}$,
\begin{eqnarray}
  \label{Eb-surface}
&&E_{\rm b}(\beta,\gamma;\epsilon_d,\kappa,\chi;N) =
  5\kappa\, N + \tfrac{N\beta^2}{1+\beta^2} 
\left[\epsilon_d + \kappa (\chi^2-4)\right]
\nonumber\\
&&\quad
+ \tfrac{N(N-1)\beta^2}{(1+\beta^2)^2}\kappa
\left[4 - 4\bar{\chi}\beta\,\cos3\gamma
  + \bar\chi^2\beta^2\right] ,\qquad
\end{eqnarray} 
where $\bar\chi\!=\!\sqrt{\frac{2}{7}}\chi$.
Explicit expressions of $g_{m_1,m_2}(\beta,\gamma)$
are given in Eqs.~(10)-(14) of~\cite{Petrellis2011}.
Diagonalization of
the matrix splits into
two (doubly degenerate) pieces with
$m=j,j-2,\ldots, -(j-1)$ and similarly with $m\to -m$.
The dimension of the basis is thus $(j+1/2)$.
The resulting eigenvalues are the
Bose-Fermi potential surfaces
$E^{(N)}_k(\beta,\gamma;\xi)$,
which include the contribution of the core and of the
single particle levels in the deformed $\beta$ and
$\gamma$ field generated by the bosons.

For $\gamma=0$ (axial shape),
the potential matrix~$\Omega_{N}(\beta;\xi)$
is diagonal in the basis $\ket{j,K;\beta;N}$
with entries~\cite{Lev88},
\bsub
\ba
&&E^{(N)}_{K}(\beta;\xi) =
E_{\rm b}(\beta;N) +
\epsilon_j + N\lambda_{K}(\beta) ~,
\label{EK-N}\\
&&\lambda_{K}(\beta;\chi,A,\Gamma,\Lambda) =
A\,\tfrac{\beta^2}{1+\beta^2}
\nonumber\\
&&
\quad
+ \tfrac{\beta}{1+\beta^2}C_{jK}\,
[\, \Gamma\sqrt{5}(\beta\bar{\chi}-2) 
  -\beta\Lambda(2j+1)C_{jK}\,] ~,\qquad
\label{tl-lamb}
\ea
\esub
where
$C_{jK} = \tfrac{3K^2-j(j+1)}
{\sqrt{(2j-1)(2j+1)(2j+3)j(j+1)}}$.
The dependence of $\lambda_{K}(\beta)$ on $K^2$
reflects the double degeneracy ($K\to -K$)
mentioned above.

The IBFM-CM~\cite{GavLevIac22b,Gav23} is an extension
of the IBFM~\cite{IBFMBook} to include
configurations with $N,N\!+\!2,\dots$ bosons,
representing shell model spaces with
0p-0h, 2p-2h,$\ldots$ particle-hole
excitations across closed shells. For two
configurations ($A,B$), the IBFM-CM Hamiltonian
can be cast in matrix form,
\ba
\hat{H}(\xi_A,\xi_B,\omega) =
\begin{bmatrix}
  \hat{H}_A(\xi_A) & \hat{W}(\omega)\\
  \hat{W}(\omega) & \hat{H}_B(\xi_B)
\end{bmatrix} ~.
\label{Hibfm-cm}
\ea
Here $\hat{H}_A(\xi_A)$ and $\hat{H}_B(\xi_B)$
represent, respectively, the normal
$A$~configuration ($N$ boson space)
and the intruder $B$~configuration ($N+2$ boson space),
both coupled to a single $j$-fermion.
Specifically,
$\hat{H}_A(\xi_A)$ and $\hat{H}_B(\xi_B)$
have the same form as in Eq.~(\ref{Hibfm}),
with a~rotational term and energy off-set,
  $\kappa^{\prime}_{B}\, \hat L \cdot \hat L + \Delta_B$, 
  added to $\hat{H}_B(\xi_B)$.
  The mixing term is
  $\hat W(\omega) =
    \omega\, [\,(d^\dag d^\dag)^{(0)} + (s^\dag)^2\,
     + {\rm H.c.} ]$,
where H.c. stands for Hermitian conjugate. 

A geometric interpretation for the IBFM-CM is obtained
by constructing from the Hamiltonian~(\ref{Hibfm-cm})
an enlarged potential matrix of order 
$(2j+1)\times(2j+1)$ in the basis
$\{\ket{j,m_1;\beta,\gamma;N},
\ket{j,m_2;\beta,\gamma;N+2}\}$,
with $m_1,m_2 = j,j~-~2,\ldots, -(j-1)$,
\ba
\Omega_{N}(y;\xi_A,\xi_B,\omega) =
\left [\begin{array}{c|c}
    \Omega_A(y;\xi_A) &
    \Omega_{AB}(y;\omega)\\
\hline
\Omega_{AB}(y;\omega) &
\Omega_B(y;\xi_B)
\end{array}
\right ] ~.
\qquad
\label{Omega-cm}
\ea
The matrix depends on $y\equiv(\beta,\gamma)$ and on
the parameters of the Hamiltonian, Eq.~(\ref{Hibfm-cm}), 
$\xi_A=(\epsilon^{A}_d,\kappa_{A},\chi,
\epsilon^{A}_j,A,\Gamma,\Lambda)$,
$\xi_B=(\epsilon^{B}_d,\kappa_{B},\chi,
\kappa^{\prime}_{B},\Delta_B,
\epsilon^{B}_j,A,\Gamma,\Lambda)$, $\omega$ and $N$.
The sub-matrix
$\Omega_A(\beta,\gamma;\xi_A)$ acts in the
$\ket{j,m_1;y;N}$ sector, 
$\Omega_B(\beta,\gamma;\xi_B)$ acts in the 
$\ket{j,m_2;y;N+2}$ sector and
$\Omega_{AB}(\beta,\gamma;\omega)$ connects the two
sectors. For $\hat{H}_{\rm f}=\hat{V}_{\rm bf} =0$,  
the above potential matrix reduces to that proposed
for the IBM-CM~\cite{Vargas04}.
Diagonalization of the matrix
$\Omega_N(y;\xi_A,\xi_B,\omega)$ produces
the eigen-potentials.

For $\gamma\!=\! 0$,
the matrix $\Omega_N(\beta;\xi_A,\xi_B,\omega)$
of Eq.~(\ref{Omega-cm}), can be
transformed into a simple block-diagonal form
$\{M_{K=j}(\beta),\,M_{K=j-2}(\beta),\,\ldots,\,
M_{K=-(j-1)}(\beta)\}$, where
$M_K(\beta)$ stands for a $2\times 2$ matrix in the
states $\ket{\Psi_{A;K}}\equiv\ket{j,K;\beta;N}$ and 
$\ket{\Psi_{B;K}}\equiv\ket{j,K;\beta;N+2}$,
\ba
M_K(\beta) =
\left [
\begin{array}{cc}
E_{A;K}(\beta) & W(\beta) \\
W(\beta) & E_{B;K}(\beta)
\end{array}
\right ] ~.
\label{MK}
\ea
Here $E_{A;K}(\beta)\!=\!
E^{(N)}_{K}(\beta;\xi_A)$, Eq.~(\ref{EK-N}),
$E_{B;K}(\beta)
\!=\!E^{(N+2)}_{K}(\beta;\xi_B)
+ 6\kappa^{\prime}_{B}\,\tfrac{(N+2)\beta^2}{1+\beta^2}
+ \Delta_B$ and
$W(\beta) \!=\!
\tfrac{\sqrt{(N+2)(N+1)}}{1+\beta^2}\omega\,
( 1 + \tfrac{1}{\sqrt{5}}\beta^2)$.

Introducing the quantities,
\bsub
\ba
\delta_{K}(\beta) &=&
E_{B;K}(\beta) - E_{A;K}(\beta) ~,
\label{delK}\\
R_{K}(\beta)
&=& \tfrac{\delta_{K}(\beta)}{2W(\beta)} ~,
\label{RK}
\ea
\label{delK-RK}
\esub
the eigen-potentials then read
\ba
E_{\pm,K}(\beta) &=&
\tfrac{1}{2}\Sigma_{K}(\beta)
\pm \big|W(\beta)\big|\sqrt{1+[R_{K}(\beta)]^2} ~,\quad
\label{EpmK}
\ea
where $\Sigma_{K}(\beta) = E_{A;K}(\beta) + E_{B;K}(\beta)$,
and the corresponding eigenvectors are
\bsub
\ba
\ket{\Psi_{-,K}(\beta)} &=& a\,\ket{\Psi_{A;K}}
+ b\, \ket{\Psi_{B;K}} ~,
\label{Psi-m}
\\
\ket{\Psi_{+,K}(\beta)} &=& -b\,\ket{\Psi_{A;K}}
+ a\, \ket{\Psi_{B;K}} ~. 
\label{Psi-p}
\ea
\label{Psi-pm}
\esub
The mixing amplitudes and probabilities satisfy
\bsub
\ba
&&\frac{b}{a} = \left [ R_{K}(\beta)
  \mp \sqrt{1+[R_{K}(\beta)]^2}\right ] ~,\\
&&b^2 = 1 - a^2 =\tfrac{1}{1+ \Bigl [ \,R_{K}(\beta)\,
    \pm \sqrt{1+[R_{K}(\beta)]^2}\,\Bigr ]^2} ~,
\label{b-prob}
\ea
\label{a-b-prob}
\esub
where the upper (lower)
sign applies for $W(\beta)>0$
[$W(\beta)<0]$.

The ensemble of
eigen-potentials~$\{E_{-,K}(\beta),\,E_{+,K}(\beta)\}$,
Eq.~(\ref{EpmK}), portray the change in energies of the odd
fermion as a function of deformation $\beta$, 
in the presence of coupled bosonic cores.
The states
$\{\Psi_{-,K}(\beta),\,\Psi_{+,K}(\beta)\}$,
Eq.~(\ref{Psi-pm}), depict the change in
configuration content.
The value of $R_{K}(\beta)$, Eq.~(\ref{RK}),
determines the character of the
normal-intruder mixing. We observe maximal mixing for
$R_K(\beta)\!=\!0$, strong mixing for
$|R_K(\beta)|\!<<\!1$,
weak mixing for $|R_K(\beta)|\!>>\!1$ and no mixing for
$|R_K(\beta)|\!=\!\infty$.

Two scenarios are relevant for the subsequent discussion.
(a)~Maximal mixing occurs for $\delta_{K}(\beta)\!=\!0$,
{\it i.e.}, when the two unmixed surfaces are degenerate,
$E_{A;K}(\beta)=E_{B;K}(\beta)\equiv E^{(0)}_K(\beta)$.
In this case, the eigen-potentials are
$E_{\pm}(\beta)= E^{(0)}_K(\beta) \pm |W(\beta)|$ and the
corresponding eigenfunctions, Eq.~(\ref{Psi-pm}),
have $\tfrac{b}{a} = +1$ ($-1$) for
$W(\beta)<0$ [$W(\beta)>0$]. 
(b)~No mixing occurs for $W(\beta)=0$.
  In this case, for $\delta_K(\beta)>0$:
  $E_{-,K}(\beta) = E_{A;K}(\beta)$,
  $\ket{\Psi_{-,K}}=\ket{\Psi_{A;K}}$, 
  $E_{+,K}(\beta) = E_{B;K}(\beta)$,
  $\ket{\Psi_{+,K}}=\ket{\Psi_{B;K}}$.
  In contrast, for $\delta_K(\beta)<0$:
  $E_{-,K}(\beta) = E_{B;K}(\beta)$,
 $\ket{\Psi_{-,K}} = \ket{\Psi_{B;K}}$, 
 $E_{+,K}(\beta) = E_{A;K}(\beta)$,
 $\ket{\Psi_{+,K}} = -\ket{\Psi_{A,K}}$. 
\begin{figure}[t]
\label{Fig1}
  \begin{center}
\begin{overpic}[width=0.51\linewidth, trim=0mm 3mm 3mm 
    0mm, clip]{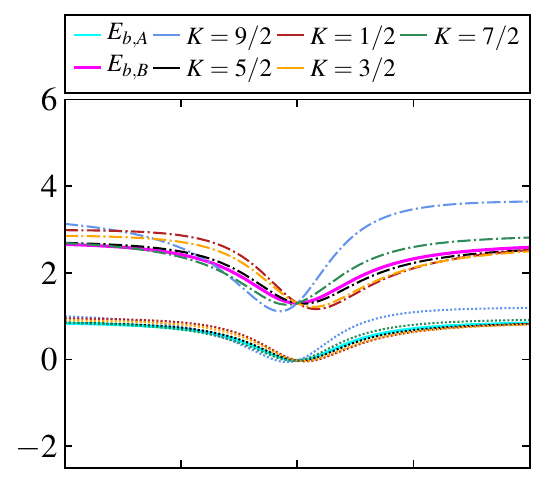}
\put(80,4) {$^{93}$Nb}
\put(15,4) {(a) $E_{A,B;K}(\beta)$}
\end{overpic}
\begin{overpic}[width=0.47\linewidth, trim=10mm 3mm 0mm 
    0mm, clip]{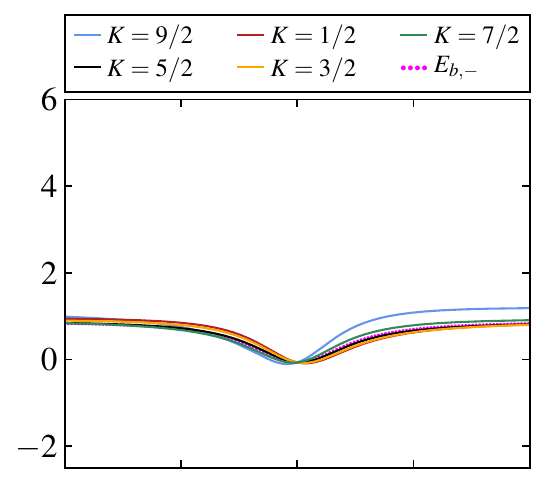}
\put(75,4) {$^{93}$Nb}
\put(5,4) {(b) $E_{-,K}(\beta)$}
\end{overpic}\\
\begin{overpic}[width=0.51\linewidth, trim=0mm 3mm 3mm 
    2mm, clip]{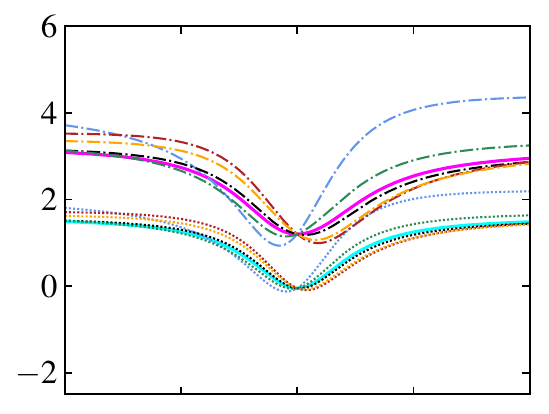}
\put(80,4) {$^{95}$Nb}
\put(15,4) {(c) $E_{A,B;K}(\beta)$}
\end{overpic}
\begin{overpic}[width=0.47\linewidth, trim=10mm 3mm 0mm 
    2mm, clip]{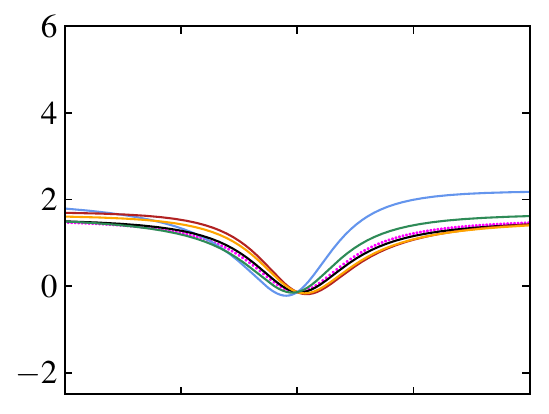}
\put(75,4) {$^{95}$Nb}
\put(5,4) {(d) $E_{-,K}(\beta)$}
\end{overpic}\\
\begin{overpic}[width=0.51\linewidth, trim=0mm 3mm 3mm 
    2mm, clip]{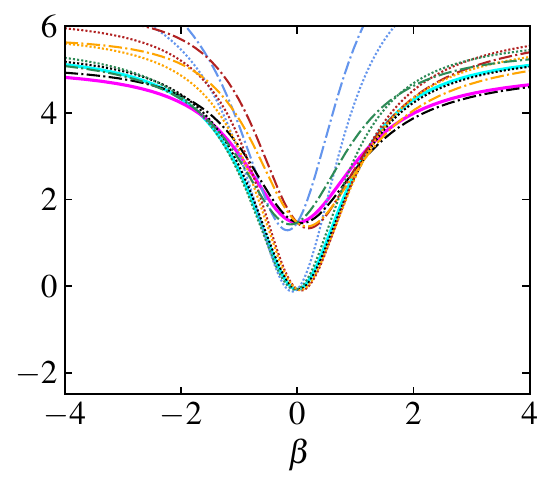}
\put(80,18) {$^{97}$Nb}
\put(15,18) {(e) $E_{A,B;K}(\beta)$}
\end{overpic}
\begin{overpic}[width=0.47\linewidth, trim=10mm 3mm 0mm 
    2mm, clip]{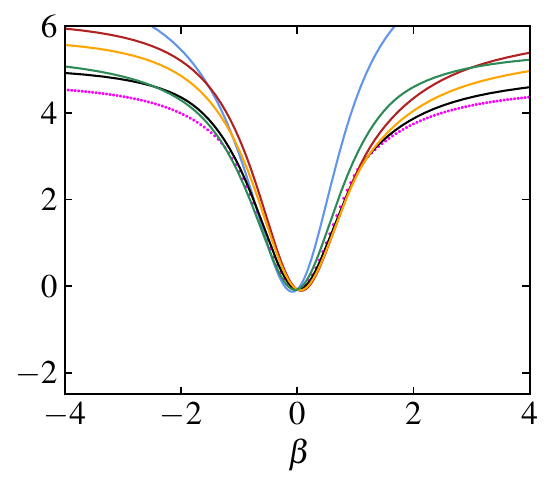}
\put(75,20) {$^{97}$Nb}
\put(5,20) {(f) $E_{-,K}(\beta)$}
\end{overpic}
\caption{\small
  Eigen-potentials $E_{-,K}(\beta)$ for each $K$,
  Eq.~(\ref{EpmK}), and unmixed surfaces
  $E_{A,B;K}\equiv\{E_{A;K}(\beta),E_{B;K}(\beta)\}$,
  Eq.~(\ref{MK}),  
in MeV for $^{93,95,97}$Nb. Purely bosonic surfaces, 
$E_{\rm b,-}(\beta)$ and
$\{E_{\rm b,A}(\beta),\,E_{\rm b,B}(\beta)\}$,
obtained by setting $\hat{H}_{\rm f}=\hat{V}_{\rm bf} =0$
in the Hamiltonian~(\ref{Hibfm-cm}), are also shown.
}
\end{center}
\end{figure}
\begin{figure}[t]
\label{Fig2}
\begin{center}
\vspace{1mm}
\begin{overpic}[width=0.51\linewidth, trim=0mm 3mm 3mm 
    2mm, clip]{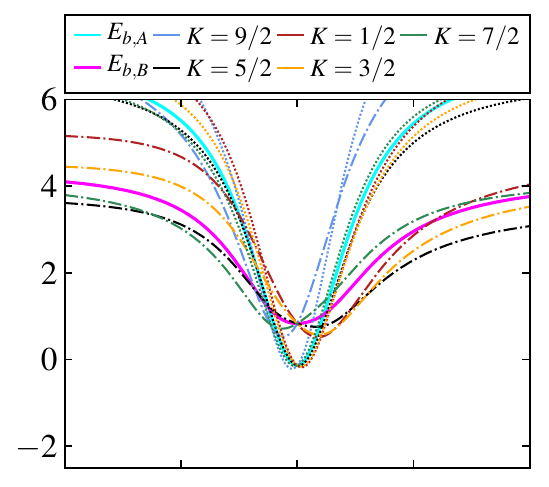}
\put(78,4) {$^{99}$Nb}
\put(15,4) {(a) $E_{A,B;K}(\beta)$}
\end{overpic}
\begin{overpic}[width=0.47\linewidth, trim=10mm 3mm 0mm 
    2mm, clip]{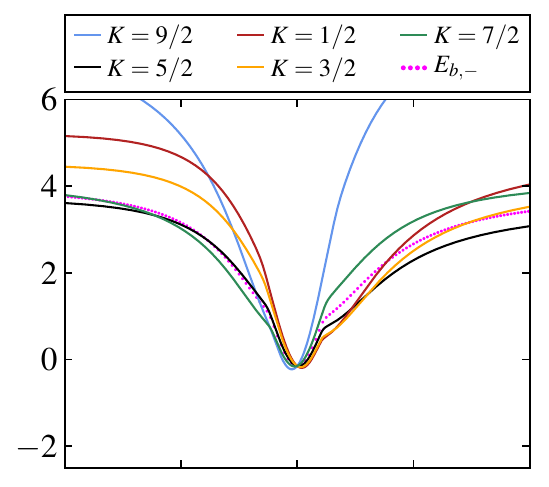}
\put(73,4) {$^{99}$Nb}
\put(5,4) {(b) $E_{-,K}(\beta)$}
\end{overpic}\\
\begin{overpic}[width=0.51\linewidth, trim=0mm 3mm 3mm 
    2mm, clip]{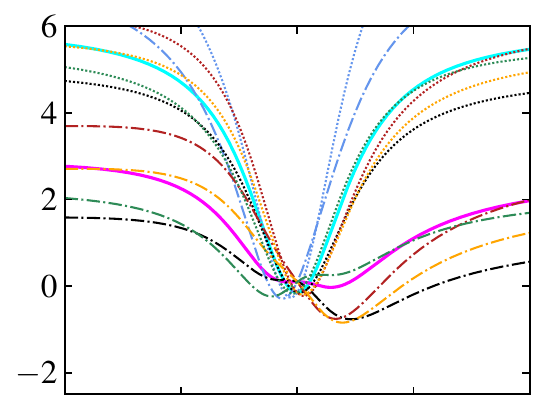}
\put(78,4) {$^{101}$Nb}
\put(15,4) {(c) $E_{A,B;K}(\beta)$}
\end{overpic}
\begin{overpic}[width=0.47\linewidth, trim=10mm 3mm 0mm 
    2mm, clip]{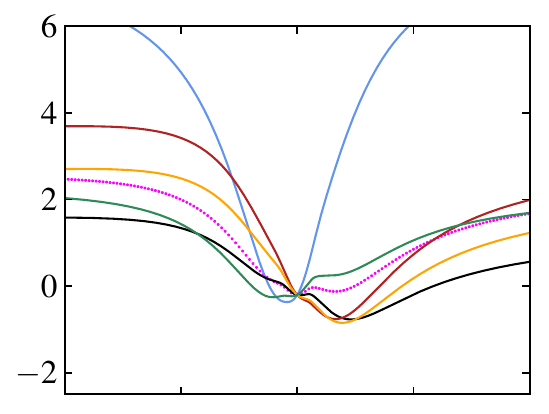}
\put(73,4) {$^{101}$Nb}
\put(5,4) {(d) $E_{-,K}(\beta)$}
\end{overpic}\\
\begin{overpic}[width=0.51\linewidth, trim=0mm 3mm 3mm 
    2mm, clip]{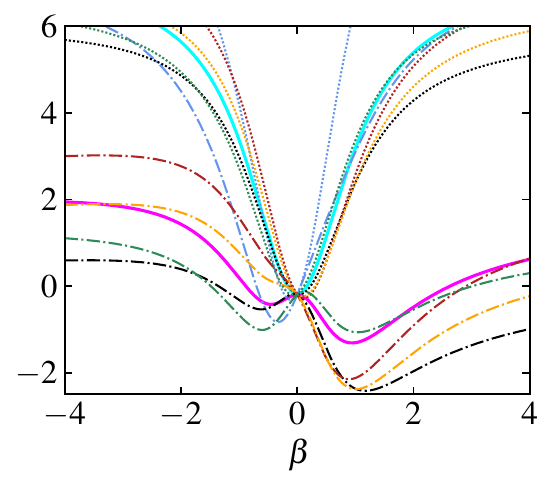}
\put(78,57) {$^{103}$Nb}
\put(14,18) {(e) $E_{A,B;K}(\beta)$}
\end{overpic}
\begin{overpic}[width=0.47\linewidth, trim=10mm 3mm 0mm 
    2mm, clip]{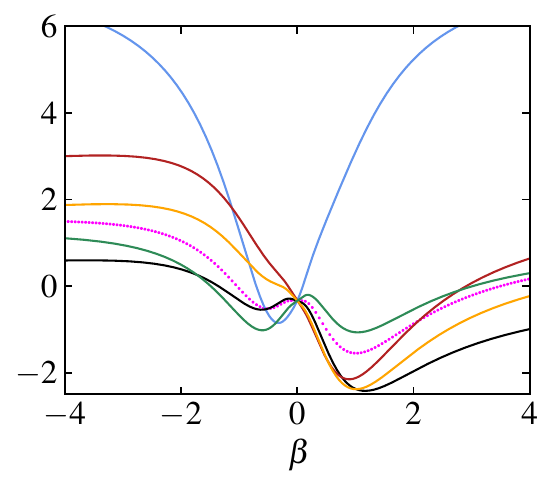}
\put(73,60) {$^{103}$Nb}
\put(5,19) {(f) $E_{-,K}(\beta)$}
\end{overpic}
\caption{\small
Eigen-potentials $E_{-,K}(\beta)$, Eq.~(\ref{EpmK}),
and unmixed surfaces
$E_{A;K}(\beta)$ (dotted lines) and
$E_{B;K}(\beta)$ (dashed-dotted lines),
Eq.~(\ref{MK}), in MeV
for $^{99,101,103}$Nb. Purely bosonic surfaces,
obtained by setting $\hat{H}_{\rm f}=\hat{V}_{\rm bf} =0$
in the Hamiltonian~(\ref{Hibfm-cm}), are also shown.
}
\end{center}
\end{figure}
\begin{figure*}[t]
\label{Fig3}
\begin{center}
  \begin{overpic}[width=0.24\linewidth]{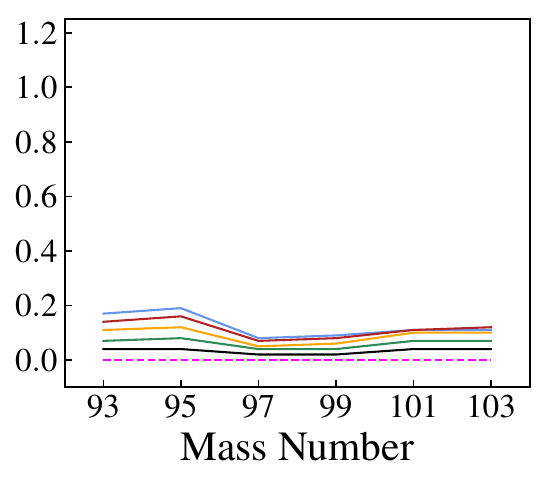}
\put(16,75) {(a) $\beta_{\rm eq}(E_{A;K})$}
\end{overpic}
  \begin{overpic}[width=0.24\linewidth]{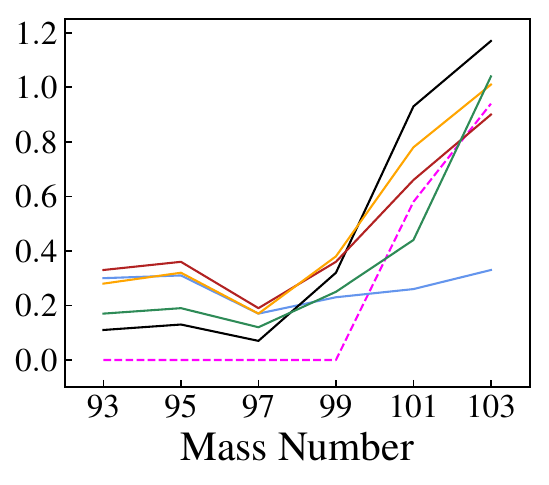}
\put(16,75) {(b) $\beta_{\rm eq}(E_{B;K})$}
\end{overpic}
  \begin{overpic}[width=0.24\linewidth]{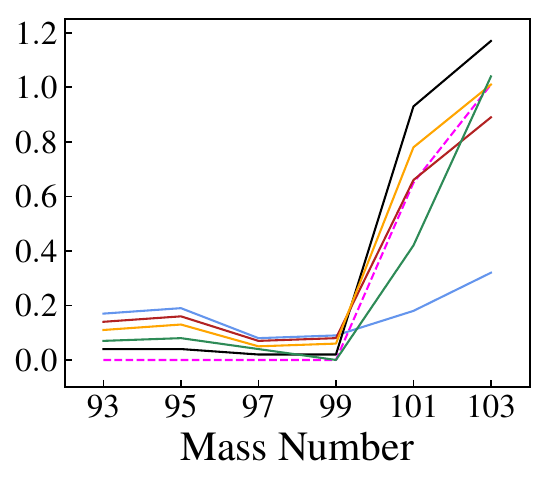}
\put(16,75) {(c) $\beta_{\rm eq}(E_{-,K})$}
\end{overpic}
  \begin{overpic}[width=0.24\linewidth]{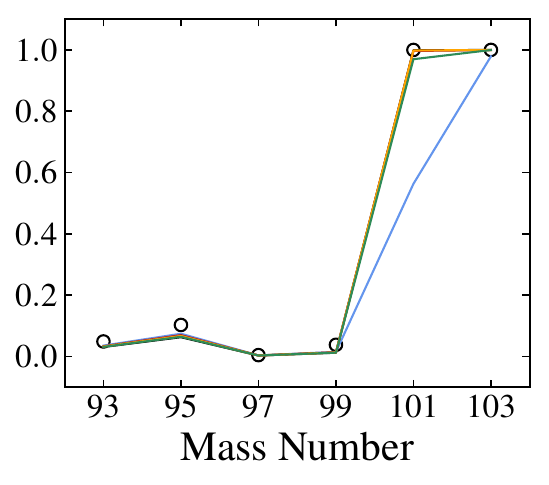}
\put(16,75) {(d) $b^2(K)$}
\end{overpic}
\caption{\small
  Equilibrium deformations (absolute value) of the
  unmixed surfaces
  (a)~$\beta_{\rm eq}(E_{A;K})$ and
  (b)~$\beta_{\rm eq}(E_{B;K})$,
  and of the lowest eigen-potential
  (c)~$\beta_{\rm eq}(E_{-,K})$.
  Dashed lines denote the minima,
  $\beta_{\rm eq}(E_{b,A}),\,\beta_{\rm eq}(E_{b,B}),
  \,\beta_{\rm eq}(E_{b,-})$, of the corresponding
  purely bosonic surfaces.
  (d)~Probability $b^2(K)$ of the intruder
  component in $\ket{\Psi_{-,K}(\beta)}$,
  Eq.~(\ref{Psi-m}), at $\beta_{\rm eq}(E_{-,K})$.
  Open circles ($\circ$)
  denote $b^2(J^{+}_{\rm gs})$ for the ground state in
  the quantum analysis~\cite{GavLevIac22b,Gav23},
  where $J^{+}_{\rm gs}\!=\!9/2^{+}$ ($5/2^{+}$)
for $^{93,95,97,99}$Nb ($^{101,103}$Nb).
Color coding for the different $K$ values as in
Figs.~1-2.}
\end{center}
\end{figure*}

The above formalism can provide
insight on the effect of configuration mixing in
Bose-Fermi systems undergoing quantum phase
transitions (QPTs).
The latter are structural changes induced by
a variation of parameters in the
Hamiltonian~\cite{Gilmore79,GilmoreFeng78},
a topic of great interest in a variety of
fields~\cite{Carr}.
As a concrete example, we apply
the formalism to the results of
a recent IBFM-CM study~\cite{GavLevIac22b,Gav23}
of QPTs in the odd-mass
Nb isotopes (Z=41) with mass number A=93-103.
Focusing on the positive parity states, the
IBFM-CM model space
$\left ([N]\oplus[N+2]\right )\otimes \pi(1g_{9/2})$,
represents in the shell model a normal
$A$~configuration of even-even Zr cores (Z=40)
and proton 2p-2h core-excited intruder
$B$~configuration, both coupled to a proton in
a~$\pi(1g_{9/2})$ orbital.
The IBFM-CM Hamiltonian employed
is that of Eq.~(\ref{Hibfm-cm}), with
parameters listed in Table~1 of the Appendix.
The quantum analysis reveals a Type~I QPT
(gradual evolution from spherical- to deformed core
shapes within the intruder configuration) superimposed
on a Type~II QPT (abrupt crossing of normal and
intruder states). The pronounced presence of both
types of QPTs demonstrates the occurrence
of intertwined QPTs in the odd-mass Nb
isotopes~\cite{GavLevIac22b,Gav23}
and in their Zr cores~\cite{GavLevIac19,GavLevIac22a}.
In what follows, we elaborate on the geometric
attributes of such structural changes.

Intertwined QPTs are characterized by a weak coupling
[small $W(\beta)$, Eq.~(\ref{MK})] and a rapid
crossing of the two configurations. At the crossing point
[$\delta_{K}(\beta)\!=\!0$, Eq.~(\ref{delK})],
the mixing in $\ket{\Psi_{\pm,K}}$, Eq.~(\ref{Psi-pm}),
is maximal.
The crossing of $E_{A;K}(\beta)$ and $E_{B;K}(\beta)$
implies a change in sign of $\delta_K(\beta)$
and the two eigenfunctions
$\{\Psi_{-,K},\, \Psi_{+,K}\}$
interchange their character. Away from the crossing point,
the system rapidly converges to the no-mixing scenario
mentioned above.

Figures 1-2 show
the lowest eigen-potentials,
$E_{-,K}(\beta)$,
for each $K$ (right panels),
which serve as the Landau potentials,
and the unmixed surfaces,
$E_{A,B;K}\equiv\{E_{A;K}(\beta),\,E_{B;K}(\beta)\}$
(left panels),
for the Nb isotopes considered.
Also shown are the purely bosonic surfaces
$E_{\rm b,-}(\beta)$ and
$\{E_{\rm b,A}(\beta),\,E_{\rm b,B}(\beta)\}$, obtained by
  taking $\hat{H}_{\rm f}\!=\!\hat{V}_{\rm bf}\!=\!0$ in the
  Hamiltonian~(\ref{Hibfm-cm}).
  For $\beta\!=\!0$, $E_{A,B;K}(\beta)$
  are independent of $K$.
  For $\beta\!\neq\! 0$, they
  exhibit quadratic and quartic $K$-splitting,
  Eq.~(\ref{tl-lamb}). The resulting landscape
  is asymmetric in $\beta$ and
is identical to that encountered in the IBFM with
a single configuration~\cite{Lev88}.
As discussed below, the presence or absence of
crossing of the unmixed surfaces, has
a direct impact on the
topology of the eigen-potentials
$E_{\pm,K}(\beta)$, Eq.~(\ref{EpmK}).

In $^{93,95}$Nb, we see from Figs.~1(a) and 1(c), that
$\delta_{K}(\beta)\!>\!0$ for all values of $\beta$,
hence for each $K$,
$E_{A;K}(\beta)$ and $E_{B;K}(\beta)$,
are well separated and do not intersect.
Consequently, the eigen-potentials, 
shown in Figs.~1(b) and 1(d),
are similar to the
unmixed surfaces,
$E_{-,K}(\beta) \approx E_{A;K}(\beta),\,
E_{+,K}(\beta) \approx E_{B;K}(\beta)$.

In $^{93}$Nb,
all $K$-surfaces $E_{-,K}(\beta)$ of Fig.~1(b), are
close in energy and are similar to the boson surface,
$E_{\rm b,-}(\beta)$, with a minimum at $\beta\!=\!0$.
This behavior reflects a spherical core shape weakly
coupled to a $j$-fermion, consistent with
the quantum analysis~\cite{GavLevIac22b,Gav23}.
The latter assigns a weak-coupling type of
wave function $\ket{(L\otimes j)J}$
with $L\!=\!0,\,j\!=\!J\!=\!9/2$,
to the normal ground state.

In $^{95}$Nb,
$E_{-,K}(\beta)$ of Fig.~1(d), display similar
topology but are more dispersed.
For small $\beta$, the
observed $K$-splitting is linear in $\beta$ and
quadratic in $K$,
in accord with Eq.~(\ref{tl-lamb}).
The factor $C_{jK}\propto [3K^2-j(j+1)]$ implies
opposite shifts for $K=1/2,\,3/2,\,5/2$ and
$K=7/2,\,9/2$ levels, with respect
to the boson surface $E_{\rm b,-}(\beta)$.

In $^{97,99,101,103}$Nb, we see from
Figs.~1(e), 2(a), 2(c), 2(e),
the occurrence of regions in $\beta$
with different signs for $\delta_{K}(\beta)$,
Eq.~(\ref{delK}),
due to crossing of the unmixed surfaces.
The crossing points
are on the prolate side at $\beta^{*}_K>0$ and
on the oblate side at $\beta^{**}_K<~0$. At these points,  
$\delta_{K}(\beta^{*}_K)=\delta_{K}(\beta^{**}_K)=0$
and $\ket{\Psi_{\pm,K}}$, Eq.~(\ref{Psi-pm}),
exhibit maximal mixing.
$\beta^{**}_K$ and $\beta^{*}_K$
mark the borders of regions with
alternating sign of $\delta_{K}(\beta)$.
I)~$\beta < \beta^{**}_K$ ($\delta_{K}(\beta) < 0$);
II)~$\beta^{**}_K < \beta < \beta^{*}_K$ 
($\delta_{K}(\beta) > 0$);
III)~$\beta > \beta^{*}_K$
($\delta_{K}(\beta) < 0$).
Region~II includes $\beta=0$.
As the system evolves from one region to adjacent~one, 
the two eigen-potentials switch their
configuration-content ($A\!\leftrightarrow\! B$).
Specifically, in regions~I and III, we find
$E_{-,K}(\beta) \approx E_{B;K}(\beta)$ and
$E_{+,K}(\beta) \approx E_{A;K}(\beta)$,
while in region~II,
$E_{-,K}(\beta) \approx E_{A;K}(\beta)$ and
$E_{+,K}(\beta) \approx  E_{B;K}(\beta)$.

In $^{97}$Nb,
the crossing of 
$E_{A;K}(\beta)$ and $E_{B;K}(\beta)$,
shown in Fig.~1(e),
occurs at high energies and their slopes
at the crossing points ($\beta^{*}_K$ and $\beta^{**}_K$)
are similar.
Consequently, these crossings have little effect on the
eigen-potentials, shown in Fig.~1(f).
In particular, the minimum of
$E_{-,K}(\beta)$ remains at $\beta=0$,
and in its vicinity $E_{-,K}(\beta)\approx E_{A;K}(\beta)$.

In $^{99}$Nb, the crossing of 
unmixed surfaces, shown in Fig.~2(a),
occurs at lower energies, and their slopes
at the crossing points are different.
This leads to ``kinks'' in the eigen-potentials
of Fig.~2(b).
Particularly noticeable are the kinks in the
$K=1/2,\,3/2,\,5/2$ levels, exhibiting a downward bend
in $E_{-,K}(\beta)$ on the prolate side.
Such kinks signal the
approach to the critical point of the Type~II QPT at
neutron number~60,
where the ground state changes from the normal to the
intruder $B$~configuration. The minimum of
$E_{-,K}(\beta)\approx E_{A;K}(\beta)$
is still at $\beta=0$.

The crossing points
$\beta^{*}_K \!>\!0$ and $\beta^{**}_K\!<\!0$ of the
unmixed surfaces
become closer to each other as the mass
number increases. Consequently, region~II
shrinks and in the heavier $^{101,103}$Nb isotopes,
the eigen-potentials satisfy
$E_{-,K}(\beta)\approx E_{B;K}(\beta)$
for most values of~$\beta$.

In $^{101}$Nb, the $E_{-,K}(\beta)$ surfaces with
$K\!=\!1/2,3/2,5/2$, develop
prolate-deformed minima in region~III,
reflecting a transition to rotational-band structure of
intruder $K$-bands. As seen in Fig.~2(d),
these deformed minima are deeper than the shallow minimum
in the flat-bottomed boson surface $E_{\rm b,-}(\beta)$ of
the even-even core, and occur at different locations. 
This highlights the effect
of the odd fermion on the QPT in the vicinity
of the critical point.

In $^{103}$Nb, all surfaces satisfy
$E_{-,K}(\beta)\!\approx\!E_{B;K}(\beta)$
and support pronounced prolate-deformed minima
(except for $K\!=\!9/2$), upon which rotational $K$-bands
of intruder states are built.
The lowest bandhead in Fig.~2(f) has $K\!=\!5/2$,
in line
with the quantum analysis~\cite{GavLevIac22b,Gav23}.
The latter assigns a strong coupling
type of wave function to members of the ground band.
The surfaces 
$E_{-,K}(\beta)$ with $K\!=\!5/2,7/2$, support also
oblate-deformed local minima.

The equilibrium deformations obtained from the
global minimum of the unmixed surfaces
and lowest eigen-potential,
$\beta_{\rm eq}(E_{A;K}),\,\beta_{\rm eq}(E_{B;K})$ 
and $\beta_{\rm eq}(E_{-,K})$, serve as the order parameters
of the QPT. Their evolution along the
Nb chain is shown in Fig.~3(a), 3(b), 3(c),
along with
$\beta_{\rm eq}(E_{b,A}),\,\beta_{\rm eq}(E_{b,B}),
\,\beta_{\rm eq}(E_{b,-})$ of the corresponding
boson surfaces. (For $^{101}$Nb,
$E_{b,-}(\beta)$ exhibits close-in-energy
spherical and deformed minima).
The normal $A$~configuration remains spherical along
the Nb chain ($\beq(E_{A;K}) \!<\! 0.2$),
while the
intruder $B$~configuration changes gradually from
weakly deformed in $^{93}$Nb
($\beq(E_{B;K})\approx \textstyle{0.3\!-\!0.4}$)
to strongly deformed in $^{103}$Nb
($\beq(E_{B.K})\approx \textstyle{0.9\!-\!1.2}$
(except for $K=9/2$).
$\beq(E_{-,K})$ is similar to
$\beq(E_A)$ for $^{93,95,97,99}$Nb
and coincides with $\beq(E_{B;K})$ for $^{101,103}$Nb.
Fig.~3(d) shows the probability $b^2(K)$ of the intruder
component in the eigenvector
$\ket{\Psi_{-,K}}$, Eq.~(\ref{Psi-m})
at $\beq(E_{-,K})$, along the Nb chain.
The rapid change in structure from the normal
$A$~configuration in $^{93-99}$Nb (small $b^2$) to the
intruder $B$~configuration in $^{101,103}$Nb (large $b^2$)
is clearly evident. The values of $b^2(K)$
calculated from Eq.~(\ref{b-prob}),
agree with the exact values of $b^2(J^{+}_{\rm gs})$
in the ground state, obtained in the quantum
analysis~\cite{GavLevIac22b,Gav23}.
The combined results of Fig.~3 
confirm the scenario of intertwined QPTs in
odd-mass Nb isotopes, where a gradual
Type~I QPT of shape changes
within the intruder configuration, is superimposed
on an abrupt Type~II QPT of configuration crossing.

In summary, we have introduced an extended
matrix coherent states
formalism for extracting the geometry of
configuration mixing in Bose-Fermi systems.
An application to the IBFM-CM, shows that the
encoded classical analysis captures
essential features of the quantum analysis
of shape evolution and coexistence in
odd-mass Nb isotopes.
This opens up opportunities to gain insight
on the physics output of
general algebraic coexistence models,
encompassing degrees of freedom with different statistics,
in terms of intuitive geometric notions.

\section*{Acknowledgements}
We thank F. Iachello (Yale) for fruitful discussions.

\section*{Appendix}
For convenience, we collect in Table~1
the information on the
parameters of the Hamiltonian, Eq.~(\ref{Hibfm-cm}),
previously used in the quantum analysis of odd-mass
Nb isotopes~\cite{GavLevIac22b,Gav23}, 
in the framework of the interacting boson-fermion model
with configuration mixing~(IBFM-CM).
The choice of parameters is discussed in the Appendix of
Ref.~\cite{Gav23}.
\begin{table}[h]
\begin{center}
\resizebox{\columnwidth}{!}{%
\begin{tabular}{cccccccc}
\hline\\[-2mm]  
& $^{93}_{41}$Nb$_{52}$ $\;$ & $^{95}_{41}$Nb$_{54}$ $\;$ &
$^{97}_{41}$Nb$_{56}$ $\;$ & $^{99}_{41}$Nb$_{58}$ $\;$ &
$^{101}_{\,41}$Nb$_{60}$ $\;$ & $^{103}_{\,41}$Nb$_{62}$ $\;$ \\[1mm]
\hline\\[-2mm]
$N$ & 1 & 2 & 3 & 4 & 5 & 6\\
$\epsilon^{A}_d$ & 0.9 & 0.8 & 1.82 & 1.75 & 1.2 & 1.2 \\
$\kappa_A$ & -0.005 & -0.005  & -0.005 & -0.007 & -0.006 & -0.006 \\
$\epsilon^{B}_d$ & 0.35 & 0.37 & 0.6 & 0.45 & 0.3 & 0.15 \\
$\kappa_B$ & -0.02 & -0.02  & -0.015 & -0.02 & -0.02 & -0.025 \\
$\kappa^{\prime}_{B}$ & 0.01 & 0.01 & 0.01 & 0.01 & 0.0075 & 0.01 \\
$\Delta_B$ & 1.6 & 1.6 & 1.84 & 1.43 & 0.8 & 0.8 \\
$\chi$ & -0.6 & -0.6 & -0.6 &  -0.6 & -1.0 & -1.0 \\
$\omega$ & 0.1 & 0.1 & 0.02 & 0.02 & 0.02 & 0.02 \\
$A$ & 0 & 0 & 0 & 0.11 & 0.2 & 0.2 \\
$\Gamma$ & 0.395 $\;$ &  0.395 $\;$ & 0.395 $\;$ &
0.395 $\;$ & 0.395 $\;$ & 0.395 $\;$ \\
$\Lambda$ & 0.362 $\;$ &  0.362 $\;$ & 1.085 $\;$ &
1.085 $\;$ & 1.374 $\;$ & 1.374 \\
\hline
\end{tabular}
} 
\caption{\label{Tab-1}
\small
Parameters in MeV
of the IBFM-CM Hamiltonian, Eq.~(\ref{Hibfm-cm}),
for the $^{A}$Nb (Z=41) isotopes.
$\chi$ is dimensionless and 
$(N,N\!+\!2)$
are the number of bosons in the $(A,B)$ configurations.
The parameters of the boson part of the Hamiltonian are
taken from Table~V of~\cite{GavLevIac22a} and of the
Bose-Fermi part from~\cite{GavLevIac22b,Gav23}.}
\end{center}
\end{table}

\end{document}